# String Theory
# -
# From Physics to Metaphysics[1]

**Reiner Hedrich**[2]

Institut für Philosophie, Fakultät Humanwissenschaften und Theologie, Universität Dortmund, Emil-Figge-Strasse 50, 44227 Dortmund, Germany

Zentrum für Philosophie und Grundlagen der Wissenschaft, Justus-Liebig-Universität Giessen, Otto-Behaghel-Strasse 10 C II, 35394 Giessen, Germany

**Abstract**

Currently, string theory represents the only advanced approach to a unification of all interactions, including gravity. In spite of the more than thirty years of its existence, the sequence of metamorphosis it ran through, and the ever more increasing number of involved physicists, until now, it did not make any empirically testable predictions. Because there are no empirical data incompatible with the quantum field theoretical standard model of elementary particle physics and with general relativity, the only motivations for string theory rest in the mutual incompatibility of the standard model and of general relativity as well as in the metaphysics of the unification program of physics, aimed at a final unified theory of all interactions including gravity. But actually, it is completely unknown which physically interpretable principles could form the basis of string theory. At the moment, "string theory" is no theory at all, but rather a labyrinthic structure of mathematical procedures and intuitions which get their justification from the fact that they, at least formally, reproduce general relativity and the standard model of elementary particle physics as low energy approximations. However, there are now strong indications that string theory does not only reproduce the dynamics and symmetries of our standard model, but a plethora of different scenarios with different low energy nomologies and symmetries. String theory seems to describe not only our world, but an immense landscape of possible worlds. So far, all attempts to find a selection principle which could be motivated intratheoretically remained without success. So, recently the idea that the low energy nomology of our world, and therefore also the observable phenomenology, could be the result of an anthropic selection from a vast arena of nomologically different scenarios entered string theory. Although multiverse scenarios and anthropic selection are not only motivated by string theory, but lead also to a possible explanation for the fine tuning of the universe, they are concepts which transcend the framework defined by the epistemological and methodological rules which conventionally form the basis of physics as an empirical science.

---

[1] The present article is a revised and extended version of a talk given at the Conference "Physik seit Einstein", organized by the Deutsche Physikalische Gesellschaft, March 2005, Berlin. Research was supported by the Deutsche Forschungsgemeinschaft under the project FA 261/7-1 "Vereinheitlichung in der Physik durch den Superstring-Ansatz: Wissenschaftstheoretische und naturphilosophische Analyse". Preliminary research was carried out during my stay (Jan. - Apr. 2002) as a Visiting Fellow at the Center for Philosophy of Science of the University of Pittsburgh.
[2] Email: Reiner.Hedrich@phil.uni-giessen.de   &   hedrich@fb14.uni-dortmund.de

# 1. The Physics and the Metaphysics of Unification

Since pre-socratic philosophy of nature, a central and dominant philosophical idea belongs to the metaphysical background of our attempts to understand nature. It is the idea that nature is a unity; that it does not exhaust itself in the plurality of phenomena that it might appear to be at first glance. In form of mathematical and theoretical concretizations, this metaphysical idea of unity, together with ontological reductionism, was of eminent importance for the development of modern physics, especially for the theoretical constructs of contemporary high energy physics. Its most advanced expression can be found in the program of a nomological unification of all fundamental interactions.[3]

A more moderate variant is conceptual unification. Its goal is to make at least compatible with each other the different theoretical drafts of our intended description of nature by eliminating contradictions in their respective results and in their prevailing requirements. It would be most interesting, if it turned out that, on the most fundamental physical level, a minimal conceptual and model theoretical unification can only be achieved at the prize of an all-encompassing nomological unification.

Many successful examples of conceptual as well as of nomological unifications can be found in the history of physics. With Newtonian physics the old separation into terrestrial and celestial mechanics was overcome. With Maxwell's electrodynamics the nomological unification of electricity, magnetism and optics could be achieved. The Einsteinian theories of relativity established the compatibility of mechanics, classical electrodynamics and, finally, gravitation.

Now, with the standard model of quantum field theory we have a theoretical construct at our disposal which achieves, at least conceptually, the unification of all fundamental forces: electromagnetic, weak, and strong - with one crucial exception: gravity. A final inclusion of gravity seems to be unattainable within the framework of quantum field theory. The apparatus of quantum mechanics and quantum field theory turns out to be conceptually incompatible with the theory of general relativity.[4]

However, all systems of nature, without any exception, are subject to gravity; and as soon as we think of all systems of nature as quantum systems, the incompatibility of quantum mechanics and general relativity is a fundamental problem. It may be true that we can ignore this problem for many microscopic as well as for many macroscopic systems: for electrons, gravity can generally be considered as irrelevant, as well as quantum effects can be ignored for planets. But for black holes and for the assumed high-density state of matter at the beginning of cosmic expansion the problem gets virulent. Here the areas of relevance of the two incompatible theoretical constructs which presently form our most advanced and most fundamental apparatus of physics are overlapping. Gravitational as well as quantum effects are crucially relevant. Neither general relativity nor quantum field theory alone is sufficient for the description of the physical phenomena in this overlap area.[5]

---

[3] Cf. Weinberg (1992).

[4] A theory of quantum gravity in the sense of a direct implementation of gravity into the quantization program of quantum field theory seems to lead unavoidably to non-renormalizability, caused by the self-interaction of the gravitons, or the non-linearity of general relativity respectively. This should be no great surprise, because it will hardly be possible to quantize gravity - in general relativity identified with the metrics of dynamical Riemannian geometry - on the flat, static Minkowski background spacetime of quantum field theory. Such a procedure will certainly not lead to the desired result. The idea of a quantization of dynamical spacetime (gravity) on the static background spacetime of quantum field theory means simply a conceptual contradiction.

---

[5] So, for example, the formation of black holes can be understood, at least partially, within the context of general relativity. According to general relativity the gravitational collapse leads to a spacetime singularity. But this spacetime singularity can not be adequately described within general relativity, because the equivalence principle of general relativity is not valid for spacetime singularities; therefore, general relativity does not give a complete description of black holes. The same problem exists with regard to the postulated initial singularity of the expanding cosmos. In these cases, quantum mechanics and quantum field theory also reach their limit; they are not applicable for highly curved spacetimes. For a certain curving parameter (the famous Planck scale), gravity has the same strength as the other interactions; then it is not possible to ignore gravity in the context of a quantum field theoretical description. So, there exists no theory which would be able to describe gravitational collapses or which could explain, why (although they are predicted by general relativity) they don't happen, or why there is no spacetime singularity.

And the real problems start, if one brings general relativity and quantum field theory together to describe black holes. Then it comes to rather strange forms of contradictions, and the mutual conceptual incompatibility of general relativity and quantum field theory becomes very clear:

Black holes are according to general relativity surrounded by an event horizon. Material objects and radiation can enter the black hole, but nothing inside its event horizon can leave this region, because the gravitational pull is strong enough to hold back even radiation; the escape velocity is greater than the speed of light. Not even photons can leave a black hole. - Black holes have a mass; in the case of the Schwarzschild metrics, they have exclusively a mass. In the case of the Reissner-Nordström metrics, they have a mass and an electric charge; in case of the Kerr metrics, they have a mass and an angular momentum; and in case of the Kerr-Newman metrics, they have mass, electric charge and angular momentum. These are, according to the no-hair theorem, all the characteristics a black hole has at its disposal.

Let's restrict the argument in the following to the Reissner-Nordström metrics in which a black hole has only mass and electric charge. In the classical picture, the electric charge of a black hole becomes noticeable in form of a force exerted on an electrically charged probe outside its event horizon. In the quantum field theoretical picture, interactions are the result of the exchange of virtual interaction bosons, in case of an electric charge: virtual photons. But how can photons be exchanged between an electrically charged black hole and an electrically charged probe outside its event horizon, if no photon can leave a black hole - which can be considered a definition of a black hole? One could think, that virtual photons, mediating electrical interaction, are possibly able (in contrast to real photons, representing radiation) to leave the black hole. But why? There is no good reason and no good answer for that within our present theoretical framework. The same problem exists for the gravitational interaction, for the gravitational pull of the black hole exerted on massive objects outside its event horizon, if the gravitational force is understood as an exchange of gravitons between massive objects, as the quantum field theoretical picture in its extrapolation to gravity suggests. How could (virtual) gravitons leave a black hole at all?

There are three possible scenarios resulting from the incompatibility of our assumptions about the characteristics of a black hole, based on general relativity, and on the picture quantum field theory draws with regard to interactions:

(i) Black holes don't exist in nature. They are a theoretical artifact, demonstrating the asymptotic inadequacy of Einstein's general theory of relativity. Only a quantum theory of gravity will explain where the general relativistic predictions fail, and why.

(ii) Black holes exist, as predicted by general relativity, and they have a mass and, in some cases, an electric charge, both leading to physical effects outside the event horizon. Then, we would have to explain, how these effects are realized physically. The quantum field theoretical picture of interactions is either fundamentally wrong, or we would have to explain, why virtual photons behave completely different, with regard to black holes, from real radiation photons. Or the features of a black hole - mass, electric charge and angular momentum - would be features imprinted during its formation onto the spacetime surrounding the black hole or onto its event horizon. Then, interactions between a black hole and its environment would rather be interactions between the environment and the event horizon or even interactions within the environmental spacetime. Our present theories do not support this picture.

(iii) Black holes exist as the product of gravitational collapses, but they do not exert any effects on their environment. This is the craziest of all scenarios. For this scenario, general relativity would have to be fundamentally wrong. In contrast to the picture given by general relativity, black holes would have no physically effective features at all: no mass, no electric charge, no angular momentum, nothing. And after the formation of a black hole, there would be no spacetime curvature, because there remains no mass. (Or, the spacetime curvature has to result from other effects.) The mass and the electric charge of objects falling (casually) into a black hole would be irretrievably lost. They would simply disappear from the universe, when they pass the event horizon. Black holes would not exert any forces on Without a theory of quantum gravity a solution to these problems is unachievable. And it is string theory (more exactly: superstring theory[6]) which promises to close this gap. String theory, incomplete as it might be, is presently the most advanced attempt at an inclusion of gravity into the unification program of physics.[7] Conceptually, an all-encompassing unification seems to be feasible with it. And, if string theory should be successful, it would be the ideal rationale for the aforementioned idea that, possibly, a minimal conceptual and model-theoretical unification will only be achievable at the prize of an all-encompassing nomological unification.

But, in some ways, with string theory we have a new situation for physics: a situation in which many of the common argumentational and methodological procedures of the empirical sciences seem to become, at least partially, inadequate. There are some hints that the unification program of physics with string theory as its possibly final stage transcends the context of physics and of the empirical sciences.[8]

## 2. Superstrings

Superstring theories are supersymmetric string theories.[9] Supersymmetry is a symmetry relation between bosons (interaction quanta) and fermions (matter particles). It encompasses the Poincaré invariance of spacetime as well as the gauge symmetries of quantum field theory. With supersymmetry, string theory postulates hitherto unobserved supersymmetric partners to the particles (and quanta) of the standard model of particle physics. However, in string theory, the basic

---

massive or electrically charged objects in their environment. They would not pull any massive objects into their event horizon and increase thereby their mass. Moreover, their event horizon would mark a region causally disconnected with our universe: a region outside of our universe. Everything falling casually into the black hole, or thrown intentionally into this region, would disappear from the universe.

A decision between these scenarios will only be achievable within the context of a new theory which overcomes the incompatibility between general relativity and quantum field theory. The same is valid for some further eminent problems and questions with regard to black holes: the Bekenstein-Hawking entropy of black holes and the microstates from which it might result, the spectrum of Hawking radiation, the information loss problem and the corresponding question concerning the conservation of quantum mechanical unitarity, etc.

[6]   See below.

[7]   The main competitor to string theory for the status of an emerging theory of quantum gravity is "Loop Quantum Gravity". As an offspring of the canonical quantization program, loop quantum gravity is a non-perturbative quantum theory of gravitation, or of geometry respectively. No unification of the forces is intended with loop quantum gravity. Cf. Rovelli (2004).

With regard to the incompleteness of both approaches to quantum gravity, Craig Callender and Nick Huggett note in the introduction to their anthology *Physics meets Philosophy at the Planck Scale*:

   "*We should emphasize at the outset that currently there is no quantum theory of gravity in the sense that there is, say, a quantum theory of gauge fields. 'Quantum gravity' is merely a placeholder for whatever theory or theories eventually manage to bring together our theory of the very small, quantum mechanics, with our theory of the very large, general relativity. [...] However, there do exist many more-or-less developed approaches to the task - especially superstring theory and canonical quantum gravity.*" (Callender / Huggett (2001a) 3) constituents of matter are not any longer matter particles and interaction quanta, but one-dimensional oscillating entities: strings. Matter particles and interaction quanta are taken to be the basic, approximately massless oscillation modes of the string.

[8]   Cf. Hedrich (2002), (2002a).

[9]   A systematical introduction to string theory can be found in Polchinski (2000), (2000a), Kaku (1999), Lüst / Theisen (1989) and Green / Schwarz / Witten (1987). For recent developments, see especially Lerche (2000), Schwarz (2000), Dienes (1997) and Vafa (1997). The early development of string theory is reflected in a commented collection of original articles: Schwarz (1985). Greene (1999) gives a recommendable popular introduction.

One of these basic oscillation modes of the string corresponds to a spin-2 particle which can be identified with the graviton, the interaction quantum of gravity. Only its discovery made string theory a candidate for a unified theory of all interactions, including gravity.[10]

Forced by intratheoretical consistency requirements (Lorentz invariance, unitarity etc.), the hitherto developed formulations of superstring theory embed the dynamics of the string into a ten-dimensional spacetime.[11] But, our common spacetime has only four dimensions. One of the solutions to this problem consists in the idea that the six surplus dimensions are comfactified microscopically in form of so-called "Calabi-Yau Spaces".[12] An alternative idea consists in treating all dimensions as macroscopically extended and assuming that open strings, whose oscillation modes represent in this picture matter particles (and ourselves), are connected to Dirichlet-branes.[13] Our observable universe would be a three-dimensional D-brane developing in time within a ten-dimensional spacetime. Only gravitons, as oscillation modes of closed strings, would move freely within this ten-dimensional spacetime. But, at the moment, this is merely a speculation.

However, the serious problems of string theory are of a different kind. On the one hand, although string theory exists since more than thirty years, it does not provide the least numerical results which could be used for an empirical test of the theoretical framework.

> *"The approximate equations that string theorists currently use are not powerful enough to work out the resulting physics fully for any choice of Calabi-Yau shape. [...] precise and definitive physical conclusions, such as the mass of the electron or the strength of the weak force, require equations that are far more exact than the present approximate framework. [...] the 'natural' energy scale of string theory is the Planck energy, and it is only through extremely delicate cancellations that string theory yields vibrational patterns with masses in the vicinity of those of the known matter and force particles."* (Greene (1999) 220)

---

[10] It can also be expected that with the transition from point particles to strings the non-renormalizability of the earlier quantum field theoretical treatments of gravity should be avoidable.

[11] There are five different perturbation-theoretical formulations of ten-dimensional supersymmetric string theory. M-Theory, a non-perturbative eleven-dimensional extrapolation from these five string theories, is not much more than a research program. Cf. Duff (1996), Schwarz (1996), (1997), Sen (1998) and Banks / Fischler / Shenker / Susskind (1997). Edward Witten, one of the most prominent protagonists of string theory, wrote in 1997 with regard to M-Theory:

> *"The novelty of the last couple of years, in a nutshell, is that we have learned that the strong-coupling behavior of supersymmetric string theories and field theories is governed by a web of dualities relating different theories. When one description breaks down because a coupling parameter becomes large, another description takes over. [...] we learn that the different theories are all one. The different supertheories studied in different ways in the last generation are different manifestations of one underlying, and still mysterious, theory, sometimes called M-theory, where M stands for magic, mystery or membrane, according to taste. This theory is the candidate for superunification of the forces of nature."* (Witten (1997) 32)

According to this picture, the five superstring theories are seen as different perturbative approximations to M-Theory. This assumption is based on the duality relations, mentioned by Witten and discovered between 1990 and 1995, which establish, under certain compactification and coupling parameter transitions, structural identities between the spectra of the oscillatory states of different perturbative formulations of string theory. These dualities reinforced significantly the uniqueness idea in string theory. See Ch. 3.

[12] Cf. Greene (1997).

[13] Cf. Polchinski (1995), (1996a), Polchinski / Chaudhuri / Johnson (1996), Douglas (1996), Bachas (1997).

On the other hand, not the least idea does exist with regard to a fundamental principle which could serve as basis, as physical motivation and as model-theoretical starting point for the development of the theory. In contrast, general relativity and its spacetime structure, into which gravity is implemented, can be understood as consequences of the principle of equivalence; the dynamics described by quantum field theories is based on local gauge invariance. No such principle is known for string theory:

> *"Ironically, although superstring theory is supposed to provide a unified field theory of the Universe, the theory itself often seems like a confused jumble of folklore, random rules of thumb, and intuitions. This is because the development of superstring theory has been unlike that of any other theory [...]. Superstring theory [...] has been evolving backward for the past 30 years. It has a bizarre history [...]. [...] physicists have ever since been trying to work backward to fathom the physical principles and symmetries that underlie the theory. [...] the fundamental physical and geometrical principles that lie at the foundation of superstring theory are still unknown."* (Kaku (1999) vii f)

> *"The story of string theory is not easy to tell, because even now we do not really know what string theory is. We know a great deal about it, enough to know that it is something really marvelous. We know much about how to carry out certain kinds of calculations in string theory. Those calculations suggest that, at the very least, string theory may be part of the ultimate quantum theory of gravity. But we do not have a good definition of it, nor do we know what its fundamental principles are. (It used to be said that string theory was part of twenty-first century mathematics that had fallen by luck into our hands in the twentieth century. This does not sound quite as good now as it used to.) The problem is that we do not yet have string theory expressed in any form that could be that of a fundamental theory. What we have on paper cannot be considered to be the theory itself. What we have is no more than a long list of examples of solutions of the theory; what we do not yet have is the theory they are solutions of. It is as if we had a long list of solutions to the Einstein equations, without knowing the basic principles of general relativity or having any way to write down the actual equation that defines the theory. / Or, to take a simpler example, string theory in its present form most likely has the same relationship to its ultimate form as Kepler's astronomy had to Newton's physics."* (Smolin (2000) 149f)

## 3. Uniqueness

Although a fundamental, physically interpretable principle which should form the basis of string theory is still unknown, and although it is not even clear, if such a principle exists at all, string theorists have emphasized all the time the internal mathematical coherence of their theoretical construct, making plausible this coherence with the fact that most of the apparent mathematical alternatives for the formulation of the theory proved to be inconclusive because of the occurrence of mathematical, physical and conceptual anomalies. This postulated coherence has been used as an argument for the uniqueness of the theoretical approach.

> *"[...] the unification of the forces is accomplished in a way determined almost uniquely by the logical requirement that the theory be internally consistent."* (Green (1986) 44)

But the crucial step in this argumentation consists in taking logical and conceptual coherence (an indispensable requirement for every theory) as sufficient to establish the adequacy of a postulated theoretical description of nature.

> *"I believe that we have found the unique mathematical structure that consistently combines quantum mechanics and general relativity. So it must almost certainly be correct."* (Schwarz (1998) 2)

In this argumentation, empirical testability seems to be irrelevant. This is understandable, at least strategically, for a theoretical approach which, after more than thirty years of development, does not provide of any empirically testable results.[14] But it should not exclusively be understood as a strategic move. Moreover, at a closer look, one can see what metaphysical intuitions form the background of the uniqueness idea:

> *"In his long search for a unified theory, Einstein reflected on whether 'God could have made the Universe in a different way; that is, whether the necessity of logical simplicity leaves any freedom at all.' With this remark, Einstein articulated the nascent form of a view that is currently shared by many physicists: If there is a final theory of nature, one of the most convincing arguments in support of its particular form would be that the theory couldn't be otherwise. The ultimate theory should take the form that it does because it is the unique explanatory framework capable of describing the universe without running up against any internal inconsistencies or logical absurdities. Such a theory would declare that things are the way they are because they* have *to be that way. Any and all variations, no matter how small, lead to a theory that - like the phrase 'This sentence is a lie' - sows the seeds of its own destruction. / Establishing such inevitability in the structure of the universe would take us a long way toward coming to grips with some of the deepest questions of the ages. These questions emphasize the mystery surrounding who or what made the unnumerable choices apparently required to design our universe. Inevitability answers these questions by erasing the options. Inevitability means that, in actuality, there are no choices. Inevitability declares that the universe could not have been different. […] the pursuit of such rigidity in the laws of nature lies at the heart of the unification program in modern physics."* (Greene (1999) 283f)

So, the metaphysical intuition behind the uniqueness argument consists primarily in the idea that things are as they are, because they have to be the way they are. There are logical constraints that make nature the way it is. There is no freedom of choice. There is only one possibility. There is no contingency. Nature is the result of an all-encompassing necessity. It could not be otherwise. The world is unique.[15]

---

[14] Although the coherence-to-uniqueness-to-adequacy argument is used in string theory for the first time as a direct substitute for empirical control, it existed even before string theory. It was, though never used as a substitute for empirical control, at least partially apparent already in general relativity and in quantum field theory.
> *"Although this contingency of physical laws has been the point of view adopted by most modern scientists, there have been some attempts even in the present century to show that physical laws are unique and necessary or that certain apparently contingent general features of the world are in fact necessary."* (Cushing (1985) 33)

Cushing saw in this tendency the first indications of a possible metamorphosis with regard to our understanding of scientific methodology and the criteria of scientific justification.
> *"The general scheme remains hypothetico-deductive, but the coherence constraint becomes much tighter."* (Cushing (1985) 32)

In string theory, the coherence-to-uniqueness-to-adequacy argument now has reached finally its climax.

[15] This metaphysical intuition also existed before string theory.
> *"A goal of science, and indeed of much of human knowledge, is to account for the world as we find it. If it could be argued that the world were either* a priori *necessary or the only one possible consistent with some general principle, then an important advance would have been made in accounting for the structure of the world."* (Cushing (1985) 33)

The uniqueness intuition does not at least form the heart of the thoroughly holistic bootstrap program which its inventor, Geoffrey Chew, introduced in the sixties, at least in a truncated form, into hadron physics. Cf. Chew (1968), (1970), (1983), Gale (1974), (1975). For the bootstrap program, the only requirement for our description of nature is self-consistency. Its central assumption is that nature is determined completely and uniquely by internal coherence:
> *"In the broadest sense, bootstrap philosophy asserts that nature is as it is because this is the only possible nature consistent with itself."* (Chew (1968) 762)

But then, so the intuition, there can be only one unique, adequate, consistent, all-encompassing description of nature. The constraints leading to the uniqueness of our world would be reflected by its theoretical description. And they would finally guarantee the adequacy of this consistent, all-encompassing description.

But, if there can be only one coherent, all-encompassing, adequate description of the world, one would need only very few requirements for our fundamental physical theory: (i) logical and conceptual consistency and (ii) most advanced universality. The first completely unified, consistently formulated, universal theory we find, would be this adequate description of nature. And the provable mathematical uniqueness of this theory would only reinforce the argument.[16]

> *"If this hope is realized, then it suffices to find that one unified theory. The first fully consistent unified theory to be found will be the only one that can be found and it will thus have to be the true theory of nature. It has even been said that, because of this, physics no longer needs experimental input to progress. At the advent of string theory, this kind of talk was very common. The transition from physics as an experimental science to physics based on finding the single unified theory was even called the passage from modern to postmodern physics."* (Smolin (2005) 26f)

The fundamental problem with the uniqueness idea is that it might be false. There could be contingency in the world. The world has not necessarily to be unique. Other worlds could be possible. Then, our consistent, all-encompassing theory has not necessarily to be adequate for the description of our world. Consistency and universality would not necessarily mean adequacy. There could be more than one consistent, universal theory, some describing other possible worlds, and only one, possibly, describing our world.

---

[16] The uniqueness idea also reinforces the hope that a fundamental nomological description of nature will not need any free parameters.
*"For many decades there has been a consensus on how to solve the problems of the undetermined parameters: unify the different forces and particles by increasing the symmetry of the theory and the number of parameters will decrease. The expectation that unification reduces the number of parameters in a theory is due both to historical experience and to philosophical argument. The former is easy to understand: [...] Newton [...] Maxwell [...]. The philosophical argument is along the lines of the following: reductionism will lead to a fundamental theory, a fundamental theory will answer all possible questions and so can't have free parameters, and unification operates in the service of greater reductionism. Or perhaps: the theory that unifies everything should be able to answer all questions. So it had better be unique, otherwise there would be unanswerable questions, having to do with choosing which unified theory corresponds to nature."* (Smolin (2004) 8)
If a fundamental theory includes free, empirically adjustable parameters, the suspicion remains that it is not really fundamental, that there should be an even more fundamental theory explaining these free parameters. The only alternative is that these free parameters reflect a fundamental contingency. Our fundamental theory would describe a spectrum of possibilities, of possible worlds, and a specific description of our world would need the specification of these free parameters. Exactly this alternative is excluded by the metaphysical thesis of the uniqueness of our world. So, at least a really fundamental description of a unique world would be programmatically incompatible with free, adjustable parameters. Free parameters reflect either contingency or that our theory is not the fundamental theory. So, the idea of a unique fundamental description of nature excludes free parameters.
Also, in the same way, the distinction between theory and contingent boundary conditions is not acceptable for a unique fundamental theory. All theoretically relevant elements, not explained by a theory, show that either there is contingency, incompatible with the uniqueness intuition, or that the theory is not fundamental.
What about the nomology described by a fundamental theory and its spectrum of solutions? Finally, also this distinction is incompatible with the concept of a unique fundamental theory. The necessity of a selection between solutions, if this selection is not determined by the theory itself, leads to the same problems; there remains only the alternative between fundamental contingency and non-fundamentality of the theory - analogous to the case of free parameters or of contingent boundary conditions. So, a complete exclusion of contingency is probably a very hard job for physics. Maybe it can only be achieved within metaphysics.

However, even given the case that our world would be unique, the relation between nature and our description of it might be rather complicated. There would be no guarantee that the logical constraints decisive for the uniqueness of the world are adequately and completely reflected within the results of our scientific endeavour. Even in the case of the uniqueness of the world, there could remain an ambiguity for possible consistent and universal theories. This could be a consequence of the procedures we apply in the development of scientific models and theories. Given such an ambiguity with regard to consistent, universal theories, only empirical tests could reinforce the adequacy claim.

Also, it might be possible that we will never be able to formulate a consistent theory of utmost universality that provides an adequate description of nature. There is no guarantee to achieve this goal within the scientific procedures and the conceptual framework at our disposal. Our epistemic capacities might be insufficient to attain an adequate, all-encompassing understanding of nature. Maybe there remains always a residuum of unexplainability, inaccessible to our epistemic means.[17] Then, an all-encompassing theory of nature would be impossible.

So, independently of postulated or proved theoretical uniqueness, of logical and conceptual consistency, and of coherence, we should not rely on metaphysical uniqueness intuitions, unless we find better arguments for these intuitions. Within the empirical sciences, uniqueness- and coherence-based arguments can not substitute empirical control. The crucial problem with mathematical models which claim to be physical theories or even adequate descriptions of nature, without having established any connection to empirical data and without even the slightest prospect of empirical control, is that the interpretation of their theoretical statements with regard to possible correlates in nature gets completely out of control.

But, as a result of recent developments in string theory, these warnings are no longer necessary. During the last years a demythologization of the uniqueness argument shook the string community. String theories seem to describe, instead of one unique world, a plethora of possible worlds. And, at the moment, it is not even clear, if our world is one of them.

## 4. Contingency

No one described the process of demythologization of the uniqueness idea in string theory better than Leonard Susskind:

> *"The world view shared by most physicists is that the laws of nature are uniquely described by some special action principle that completely determines the vacuum, the spectrum of elementary particles, the forces and the symmetries. Experience with quantum electrodynamics and quantum chromodynamics suggests a world with a small number of parameters and a unique ground state. For the most part, string theorists bought into this paradigm. At first it was hoped that string theory would be unique and explain the various parameters that quantum field theory left unexplained. When this turned out to be false, the belief developed that there were exactly five string theories with names like type-2a and Heterotic. This also turned out to be wrong. Instead, a*

---

[17]     Cf. Hedrich (1998), (2001), (2002b).

*continuum of theories were discovered that smoothly interpolated between the five and also included a theory called M-Theory. The language changed a little. One no longer spoke of different theories, but rather different solutions of some master theory."* (Susskind (2003) 1)

Significantly more ironic is Lee Smolin's comment:

*"[...] the number of string theories for which there is some evidence for has been growing exponentially as string theorists developed better techniques to construct them."* (Smolin (2004) 10)

One of the fundamental problems with this multitude of string scenarios is to set them in relation to observable phenomenology. So, it should, not at least, be possible to identify the quantum field theoretical standard model of elementary particle physics, with its interactions and symmetries, as one of the low energy limits resulting from the spectrum of possible string scenarios.[18] But this seems to be rather problematic. There exists in string theory an immense number of possibilities for the broken symmetries in the energy range of the standard model. In Michio Kaku's words, there are

*"[...] millions of ways to break down the theory to low energies."* (Kaku (1999) 17).

The ambiguity with regard to the resulting low energy symmetries is, not at least, a consequence of the ambiguity caused by the immense number of possibilities for the transition from the original ten-dimensional dynamics of string theory, forced by the requirement of internal consistency, to the phenomenologically relevant implications for a four-dimensional spacetime.

*"Now, we would be more than glad if strings would remain in lower dimensions as simple as they are in D = 10. However, especially string theories in D = 4 turn out to be much more complicated."* (Lerche (2000) 18)

In the geometrical picture of compactification, this ambiguity of the transition corresponds to the multitude of answers to the question in which form the six surplus dimensions can be compactified. None of the possible compactifications has a better theoretical justification than any other.

*"There is no known reason why a ten dimensional theory wants at all to compactify down to D = 4; many choices of space-time background vacua of the form $R^{10-n} \times X_n$ appear to be on equal footing."* (Lerche (2000) 19)

For compactification, there exists (i) a large number of possible Calabi-Yau spaces with different topologies. For every topology there exists additionally (ii) a continuum of geometric parameters. And (iii), for every point on the four-dimensional spacetime, resulting from the compactification of the six surplus dimensions, these parameters can be different.

*"First, there are many topologies of Calabi-Yau manifolds, which represent discrete choices for the configurations of the extra six dimensions. [...] But even worse, there are* continuous families *of Calabi-Yau*

---

[18]   In this context, the free parameters of the standard model should be reproduced and explained by a theoretical scenario which works, on the fundamental level, without these parameters. It is not at all clarified, if this will be possible:
*"All currently accepted physical theories require some phenomenological input. Our recent enthusiasm for string theory as the theory of everything has given rise to the hope that the only necessary inputs are the basic dimensionful parameters which define the conversion between socially defined scales of measurement and the fundamental units of mass, length, time and action. This is not necessarily the case, and the existence of mathematically consistent, disconnected, models of quantum gravity suggests that it* is *not the case."* (Banks / Dine / Gorbatov (2003) 21)

*manifolds, where the shape and size of the manifold varies continuously. Moreover, these parameters may vary as a function of four-dimensional coordinate x. [...] This variation is parametrized by a four-dimensional field R(x) giving the characteristic size as a function of position."* (Giddings (2005) 10)

So, there is an immense number of possible combinations of these partially continuous and partially discrete geometrical and topological parameters which characterize the structure of the corresponding compact manifold and which can vary for every point on the four-dimensional spacetime resulting from the compactification. These parameters are usually called "moduli".[19]

Most important for the relation between the spectrum of string scenarios and phenomenology is that different parameter combinations, corresponding to different compactification scenarios, lead in almost all cases to completely different physical results on the extended four-dimensional spacetime remaining after compactification. So, different compactification scenarios mean different physics: different symmetries, different particle spectra and masses, different effective low energy parameters.

*"[...] there are many Calabi-Yau three-folds and each gives rise to different physics in $M_4$. Having no means to choose which one is 'right', we lose predictive power."* (Greene (1997) 9)

And this physical and, consequently, phenomenological variability seems to be independent from the question, if the six compactified surplus dimensions of string theory should be interpreted geometrically as real space-dimensions or rather as a useful picture for an internal dynamical parameterization. It seems to be independent of the question, if compactification should be understood in a realistic sense as spacetime phenomenon or rather as a structural metaphor for the parameterizable multitude of expression modes of the low energy phenomenology of a still unknown fundamental theory of which string theory is an approximation.

Let's have a closer look at the problems resulting from the spectrum of string scenarios. Every point of the configuration space of possible moduli combinations of supersymmetric string theory, the so-called "supermoduli-space", represents in the context of the compactification picture a specific six-dimensional Calabi-Yau space and corresponds to a resulting effective low energy nomology for the extended four-dimensional spacetime: a string vacuum.[20] To compare these string vacua with the observable phenomenology, it would be necessary to derive the low energy implications (e.g. the parameters for the effective low energy quantum field theories) for all possible moduli

---

[19] *"Examples of moduli are the size and shape parameters of the compact internal space that 4-dimensional string theory always needs. [...] In a low energy approximation the moduli appear as massless scalar fields."* (Susskind (2003) 1)
Appearing as massless scalar fields within the four-dimensional spacetime, the moduli lead to long-range interactions, being in competition with gravity and thereby violating the equivalence principle. The hope remains that they are an artifact of the perturbative approach of string theory.
*"These massless fields are all called moduli fields, and they are a desaster. [...] parameters in the four-dimensional lagrangian, such as fermion masses and coupling constants will all vary with the moduli. Worse still, the modulus fields interact with the other fields of the theory with gravitational strength. Massless scalars with such interactions lead to fifth forces, time-dependent coupling constants, and/or extra light matter, none of which are seen experimentally."* (Giddings (2005) 10)

[20] For a first comparison with phenomenology, it is probably sufficient to restrict considerations to the string vacua, because they reflect the most relevant low energy implications.
*"[...] most of the physics that is observable at low energies seems to be governed by the vacuum (zero mode) structure and not by the microscopic theory, at least as far as we can see today."* (Lerche (2000) 19)

combinations. But, this can not be achieved within the framework of perturbative string theory.[21] However, even if the calculation of the low energy implications of a specific compactification scheme (or even of all compactification schemes) were possible, the problem would remain that there are a lot of different string vacua amongst which we had to look for the symmetries and the coupling parameters of the standard model, or for the phenomenologically adequate scenario respectively. And we would probably have to explain afterwards what distinguishes the identified vacuum from all the others. We would have to explain, why exactly the identified vacuum is realized in our world.

And it is a massive understatement to talk about a lot of string scenarios: recent estimations suggest between $10^{100}$ and $10^{500}$ effective four-dimensional string vacua.[22] For this spectrum recently the terms "landscape"[23] and "discretuum"[24] were introduced.[25]

The string landscape can be best understood as a multidimensional configuration space of the parameterization of possible effective physical scenarios (worlds), with different symmetries, with different interactions and interaction structures, with different coupling parameters, with different particle spectra and particle masses.

> *"For practical purposes, the landscape gives us a large set of alternative effective Lagrangians for describing the physics we have observed in our universe. These are parametrized by a collection of numbers, which include the dimensions of space-time, the name, rank and representation content of the low energy gauge theory, the value of the cosmological constant, and the values of all the coupling constants and masses of fields in the Lagrangian. These numbers can be collected together and viewed as a multidimensional probability space."*
> (Banks (2004) 16)

The connective structure within the landscape, especially the possible combinations of discrete and continuous parameters, is presently far from clarified.

> *"[...] the notion of connectedness in this landscape is, at best, obscure."* (Dine (2004a) 3)

In particular, Thomas Banks emphasizes again and again the

> *"[...] disconnected moduli space [...]."* (Banks (2003) 18)

Banks has fundamental doubts about the existence of the landscape, because the landscape hypothesis is the result of calculations based on the apparatus of effective field theory. According to Banks this is completely inadequate in the context of string theory and leads to highly questionable results.[26]

---

[21] And there does not exist any consistent non-perturbative formulation of string theory, a problem connected probably to the fact, that the fundamental physical principle of string theory is still unknown.

[22] Until now, the investigation of string vacua and their statistics has been carried out exclusively by means of approximations for weak coupling, and by subsequent combinatorial reasoning. The correctness of the results depends on the hope that the duality relations between the perturbative formulations of string theory lead to a sufficiently representative picture of the spectrum of string vacua. Cf. Douglas (2003).

[23] Cf. Susskind (2003).

[24] Cf. Banks / Dine / Gorbatov (2003).

[25] The question, if the string landscape includes sectors which do not result from the supermoduli space of supersymmetric vacua, will be discussed below.

[26] Cf. Banks (2004), (2003), Banks / Dine / Gorbatov (2003).

> *"[...] the landscape of string theory is far from an established fact. [...] my personal bottom line on this subject is that the Landscape probably does not exist."* (Banks (2004) 2)

Banks points out to the fact that there exists no common action function in the sense of quantum field theory for the different string scenarios of the landscape. So, the idea of different solutions to the same theory and, especially, of the possibility of transitions between these solutions by variation of certain parameters is, according to Banks, not applicable to string theory.

> *"Banks has argued cogently that one cannot use effective field theory to study multiple vacua in theories of gravity. For example, in many circumstances there are no transitions between the different states, and an observer in one can not do experiments which will indicate the existence of others. So it is not clear that the multiplicity of states has any meaning."* (Dine (2004) 7)

It is an illusion, as Banks suggests, that the different string vacua of the landscape are resulting from the same theory. Different moduli-combinations belong, according to Banks, to completely different Hamilton functions.

> *"The notion of different vacua of the same theory, in any of the senses that this is meant in quantum field theory, is simply not applicable to theories of quantum gravity, beyond the very limited context of continuous moduli spaces of Super-Poincaré invariant S-matrices."* (Banks (2004) 3f)

So, either we can get to the conclusion that the landscape of string theory does not exist.

> *"Still, the possibility that the landscape may not exist should be kept in mind."* (Dine (2004a) 3)

Or, to interpret Banks' arguments in another way, the landscape does not consist of different solutions of the same theory, but of an immense multitude of per se autonomous theoretical scenarios and nomological structures which belong, in a certain way, to the same family of theories. Then, transitions between these scenarios do not exist. One can not get dynamically, by variation of parameters, from one scenario to another.

The problem of the relation between the landscape of string theory and the observable low energy phenomenology remains. And even if the landscape hypothesis will be elaborated in a more conclusive way within string theory, it is evidently impossible that a vacuum which resembles our world will be found within the vacua resulting from the supermoduli-space. The crucial point of this particular problem is supersymmetry. Superstring theories can, at least in the traditional perturbative formulation, only be consistently formulated, if they include supersymmetry. Without supersymmetry, mathematical anomalies and non-renormalizable divergences are unavoidable. Consequently, all vacua resulting from the supermoduli-space of string theory are supersymmetric.[27]

---

[27] But, *"[...] the supersymmetry forbids any potential on the moduli space."* (Banks / Dine / Gorbatov (2003) 19) This means that all vacua resulting from the supermoduli-space would be energetically degenerated. There wouldn't be any energy differences between the string vacua resulting from the supermoduli-space which might help to identify and select a vacuum which could be assumed as being realized physically in our world.
And, supersymmetry has a further decisive consequence: For all string vacua resulting from the supermoduli-space the cosmological constant is necessarily zero.
> *"The value of the potential energy at the minimum is the cosmological constant for the vacuum. [...] The supermoduli-space is a special part of the landscape where the vacua are supersymmetric and the potential [...] is exactly zero. [...] On the supermoduli-space the cosmological constant is also exactly zero. Roughly speaking, the supermoduli-space is a perfectly flat plain at exactly zero altitude."* (Susskind (2003) 2)

This seems to be incompatible with recent astrophysical data and with their corresponding cosmological implications.

But our world is obviously not fully supersymmetric. Unbroken supersymmetry is incompatible with observable phenomenology. In a fully supersymmetric world, our known matter and interaction particles and their corresponding supersymmetric partners would have the same masses. If they would have the same masses, we would already have found these supersymmetric partners. So, only a broken supersymmetry can be realized in our world. Because all string vacua resulting from the supermoduli-space are fully supersymmetric, obviously none of these vacua can give a description of our world.

> *"So far, no string theory background is known which is consistent with all features of the observed universe. They all have one or more of the following features, which each disagree with observation: no positive cosmological constant, unbroken supersymmetry, massless scalar fields."* (Smolin (2003) 48)

If the spectrum of string vacua were exclusively resulting from the supermoduli-space, string theory must be wrong. Supersymmetric vacua can not reproduce observable phenomenology. So, either string theory is wrong, or the supermoduli-space can not be everything which contributes to the spectrum of string vacua.

> *"[...] the continuum of solutions in the supermoduli-space are all supersymmetric with exact super-particle degeneracy and vanishing cosmological constant. Furthermore they all have massless scalar particles, the moduli themselves. Obviously none of these vacua can possibly be our world. Therefore the string theorist must believe that there are other discrete islands lying off the coast of the supermoduli-space. [...] This view is not inconsistent with present knowledge [...] but I find it completely implausible."* (Susskind (2003) 1)

How could a broken supersymmetry result from string theory without leading to mathematical and physical anomalies? Ten-dimensional perturbative string theories are necessarily fully supersymmetric. The only possibility imagined until now consists in the idea that supersymmetry breaking is a result of compactification. The mathematics of compactification should therefore make understandable how a broken supersymmetry and the fact that we don't observe supersymmetry multiplets in our world can result from a fully supersymmetric theory. There should be compactifications which lead from a fully supersymmetric ten-dimensional theory to a broken supersymmetry for the resulting four-dimensional vacuum. Then, the string landscape would contain vacua which do not result from the supermoduli-space. But this extension of the spectrum of string vacua is actually no more than a speculative extrapolation.[28]

---

[28] > *"A key problem has been constructing string theories that agree with the astronomical evidence that the vacuum energy (or cosmological constant) is positive. The problem is that a positive cosmological constant is not consistent with supersymmetry. But supersymmetry appears to be necessary to cancel dramatic instabilities having to do with the existence of tachyons in the spectrum of string theories."* (Smolin (2004) 10)

Banks has his doubts with regard to this extrapolation.
> *"There are many disconnected continuous families of Poincaré invariant solution to string theory. They have various dimensions, low energy fields, and topologies, but they all share the property of exact [supersymmetry]. The program of string phenomenology is to find a [supersymmetry] violating, Poincaré invariant solution of the theory, which describes low energy scattering in the real world. In [Banks (2001)] I expressed the opinion that no such solution exists."* (Banks (2003) 2)

He points out that there are not even approximate solutions indicating the possibility of a broken supersymmetry.
> *"There are no known asymptotically flat string vacua with broken [supersymmetry]."* (Banks (2003) 30)
> *"As of this date, we know of no example of a controllable approximation to a theory of quantum gravity that leads to a nonsupersymmetric theory in asymptotically flat, Poincaré invariant spacetime. [...] This failure [...] leads me to conjecture that there are* no *[supersymmetry] violating, Poincaré invariant theories of quantum gravity."* (Banks (2003) 7f)

Nonetheless, some string theorists hope to be able to establish a definition for a non-zero potential for these vacua with broken supersymmetry.

## 5.     Selection

For a long time string theorists held to the idea that there can be only one consistent, fundamental theory which describes our world, and that this theory is string theory. Meanwhile, it turned out that the string approach very probably leads to an immense multitude of theoretical scenarios describing possible worlds with different effective low energy nomology.

> *"For low energy observers, physics is different in each of these states. Gauge groups, coupling constants and the like all vary. The cosmological constant, in particular, is a random variable in these $10^{1000}$ (?!) states."* (Dine (2004a) 6)

If at all, only one of these theoretical scenarios will describe our world with its specific low energy phenomenology, its interactions, its gauge invariances, its coupling parameters, its particle spectrum, its particle masses, and its spacetime structure and dimensionality.

> *"We believe that there are many mathematically consistent models of quantum gravity, at most one of which describes the real world."* (Banks / Dine / Gorbatov (2003) 5)

But, which of the theoretical scenarios the string approach leads to, which of its string vacua, corresponds to our world?

> *"Despite the unity of the theory, string/M theory appears to describe a very large number of four dimensional (and other) vacua with inequivalent physics, most of which clearly do not describe our universe. At present we have no clue which one is relevant, or how to find it."* (Douglas (2003) 1)

---

> *"[...] the supermoduli-space is a perfectly flat plain at exactly zero altitude. Once we move off the plain, supersymmetry is broken and a non-zero potential develops [...] Thus beyond the flat plain we encounter hills and valleys. We are particularly interested in the valleys where we find local minima of V."* (Susskind (2003) 2f)

But, as long as there are no reliable mathematical foundations for this picture, it is pure speculation.

> *"No perfectly precise definition exists in string theory for the moduli fields or their potential when we go away from the supermoduli-space."* (Susskind (2003) 17)

According to Banks, the idea of defining an effective potential for the landscape is not well-founded, if not meaningless at all:

> *"Much of the thinking implicit in discussions of flux compactifications depends on the notion that the effective potential is an exact object, for which we are presently able to find only approximate expressions. This line of thought might be completely wrong. We have no evidence from string theory or gravity that such an object exists."* (Banks / Dine / Gorbatov (2003) 8)

> *"In my opinion, the concept of an effective potential on moduli space as a tool for finding string models of gravity, is a snare and a delusion, fostered by wishful thinking, and without regard to the actual evidence in front of us. There is no evidence for this concept in solid string theory calculations, and lots of evidence against it."* (Banks (2004) 4)

> *"The notions of vacuum, effective potential, and vacuum decay from field theory, are not correct ones in quantum gravity."* (Banks (2003) 24)

> *"[...] it is clear that quantum gravity in asymptotically flat spacetimes is a different kind of beast from quantum field theory, with a high energy density of states unlike any quantum theory we have dealt with before. [...] Much of the conventional framework of quantum mechanics is lost."* (Banks (2003) 27)

If there exist $10^{100}$ or $10^{500}$ different string vacua, how can we identify the adequate scenario? At the time, it is completely implausible that the low energy implications for these different string scenarios could be calculated.

> *"It is clear that at a certain point in this process, if we don't falsify the landscape easily, we will run into the problem that current technology does not allow one to calculate the low energy parameters with any degree of precision. Indeed, the error estimates are only guesses because we don't even know in principle how to calculate the next term in the expansion in large fluxes."* (Banks (2004) 19)

But, even if it would be possible to derive the low energy phenomenology for these scenarios, the immense number of string vacua would lead to a problem. Clarity would only be achieved, if it could be shown that the string approach, on principle, does not lead to any scenario compatible with observable low energy phenomenology. Then string theory would simply not be able to describe our world. This would falsify string theory.

> *"String theory could fail if there turn out to be no consistent and stable string vacua consistent with all the observed features of our universe including complete supersymmetry breaking, the absence of massless scalar fields and a positive cosmological constant. / Conversely, string theory could fail if it turns out that there are so many consistent and stable string vacua consistent with all observations to date that they populate the space of post-standard model physics densely enough that the theory makes no predictions for future experiments."* (Smolin (2003) 52)

If a falsification via incompatibility with low energy phenomenology will not take place, the problem remains. If there are $10^{100}$ or $10^{500}$ string vacua, and if there are no constraints which exclude vacua resembling our world on principle, then, even after a preselection which leaves only those vacua which resemble more or less our world, there will probably still remain an immense multitude of possible vacua, compatible with the observable phenomenology, but with different nomology. And, if this preselected ensemble of vacua corresponds to a continuum or a dense discretuum of parameter values, it will be impossible to identify by empirical means the vacuum representing our world, and therefore the nomology it is based on, because the fixing of the relevant parameters by measurement will only lead to a finite exactitude.

> *"However, if one takes the possibility of the existence of these theories seriously, there is a disturbing consequence. For the number of distinct theories that the evidence points to is vast, estimates have been made on the order of $10^{100}$ to $10^{500}$. Each of these theories is consistent with the macroscopic world being four dimensional, and the existence of a positive and small vacuum energy. But they disagree about everything else, in particular, they imply different versions of elementary particle physics, with different gauge groups, spectra of fermions and scalars and different parameters. / That is, if the string theorists are right, there are on the order of $10^{100}$ or more different ways to consistently unify gauge fields, fermions and gravity. This makes it likely that string theory will never make any new, testable predictions concerning the elementary particles. / Of course, a very small proportion of these theories will be consistent with the data we have, to date, about particle physics. Suppose this is only one in $10^{50}$. There will still be $10^{50}$ different theories, which will differ on what we will see in future experiments at higher energy. This number is so vast, it appears likely that whatever is found, there will be many versions of string theory that agree with it."* (Smolin (2004) 11)

We would never get from the observable phenomenology to the nomology from which it results. There would simply be too much possible nomologies behind the observable phenomenology and compatible with it.

> *"Accepting, at least provisionally, the existence of the landscape, the nature and goals of string theory (fundamental physics) are different than we previously imagined. In this vast 'landscape', one can't hope to find 'the state' which describes our universe. Nor can our goal be to predict all of the features of nature with arbitrary precision."* (Dine (2004a) 3f)

Even if conclusive predictions could be achieved for all nomologies, empirical methods would not be sufficient to decide which vacuum and which nomology is ours. Even if the complete spectrum of its possible low energy implications could be calculated, string theory would finally have no predictive power. Even if all relevant implications for all string scenarios could be derived, string theory would, on principle, be a theory which can not be falsified. String theory would remain metaphysics, without any hope to become a physical theory.

But, even if string theory should remain a metaphysical theory, and even if it should be impossible to identify the string vacuum which is realized in our world: if string theory is the "right" metaphysical theory (how could we know?), which means, not at least, that one of its vacua corresponds to our world, the question remains: what makes this vacuum so special that it is realized? This is the so-called "vacuum selection problem" of string theory. What inherent necessity, what dynamical process, what structural constraints, what nomological implication, selects the one string vacuum that is realized from the multitude of possibilities? Or is it simply coincidence? Is our world simply a contingent selection from an immense number of possibilities? - This ontological question remains, independently of the epistemological question, if we can or can not identify this vacuum within the spectrum of possibilities by means of empirical research.[29]

> *"Even if a string theory background is found which is [...] consistent with everything that is observed, does this tell us anything, given that there is an infinite space of possible string backgrounds to search? [...] So, we should ask, even if there is a unique string theory background consistent with what is observed, how could nature pick it out? One might hope that there were a principle of stability or lowest energy that would pick out a unique string theory background. However this is unfortunately unlikely."* (Smolin (2003) 48)

The only solution to the vacuum selection problem would consist in a plausible, conclusive selection principle. But, even Michael Douglas, a specialist in the statistics of string vacua, sees no chance to find such a selection principle.

> *"[...] there is a widespread feeling that a 'theory of everything' should make unique predictions for the physics we observe. String/M theory as we understand it now does not do this, and it is this lack which is often cited as the reason why a 'Vacuum Selection Principle' should exist. Of course, this argument in itself is simply wishful thinking."* (Douglas (2003) 5)

Not the least idea seems to exist, why a specific scenario out of the spectrum of possibilities is realized with our world.

> *"The recent progress in non-perturbative string theory does not solve the problem of the choice of vacuum state either. The progress is rather conceptual [...]."* (Lerche (2000) 20)

The fundamental question behind the vacuum selection problem of string theory is, if it will ever be possible to deduce all features of the world from a fundamental description of nature, or if this strategy, afflicted with contingency as the world might be, won't be successful. Does the road to a unified and all-encompassing description of nature lead to an elimination of contingency or will our universe turn out to be only one contingently selected possibility out of a vast spectrum of consistently possible scenarios?

---

[29] Coupled to the vacuum selection problem are a great number of open questions and problems: the determination of gauge invariances, of coupling parameters, of particle spectra, families and masses, of a broken supersymmetry, etc.

Nomological extrapolation is carried as far as possible in string theory. But, meanwhile, with the landscape hypothesis and in absence of any reasonable selection principle, is seems very improbable that a complete elimination of contingency can be achieved in the context of string theory. String theory does not support the idea that our world and its specific features are necessary. Moreover, the uniqueness hypothesis to which string theorists held for a long time, and, with it, the background assumption of a contingency-free, in its features completely necessary world, the only consistent world at all, got a highly problematic status with the recent developments in string theory, not to say that it was completely brought down with the landscape hypothesis. If string theory gives an adequate description of nature (a probably unprovable assumption, if the following is correct), and if the landscape hypothesis should turn out to be string theory's inevitable consequence, that would mean that there is obviously not only one consistent structure describing the one and only consistent world, but a lot of such structures, and therefore a lot of possible worlds. Our world would simply be contingent with regard to its specific features and its structure. Many structures, or many worlds respectively, with different physics, would be possible. Ours would be only one of these possible worlds.

In this case, it might be that a unified, fundamental theory would not especially refer to our world, but rather to the whole spectrum of possibilities. Our universe (to use the language of the dominant model-theoretical apparatus of physics, i.e. systems of differential equations) would perhaps be represented by one of the many solutions of the fundamental equations of a final, unified theory. The fundamental equations of the theory would represent the whole ensemble of possible worlds. Under these conditions, a unified theory would not determine the features and the nomological structure of our world without the specification of some free parameters. These parameters would be a constitutive part of the solution representing our world. With regard to the fundamental equations of the unified theory, they would have to be considered as boundary conditions.

Without any selection principle, it is not very far from the contingency hypothesis to the ensemble hypothesis.[30] The corresponding transition consists simply in the idea, that physically possible scenarios are not only possible, but (whatever that means) physically real, like our world, forming an ensemble of real worlds: a multiverse.[31] But, naturally, to talk about an ensemble of causally

---

[30] An independent motivation for the ensemble hypothesis consists in the fine-tuning of the universe. See below.

[31] It is useful, as a first concretization of the ensemble idea, to understand the worlds within such an ensemble as causally unconnected realizations of possibilities. This implies that it would be completely meaningless to think about the worlds within the ensemble as existing parallely in a temporal sense. They would only share a common configuration space of structural and nomological possibilities.

Lee Smolin gives an explicit formulation of the ensemble hypothesis with regard to quantum gravity:

*"**A** There exists (in the same sense that our chairs, tables and our universe exists) a very large ensemble of 'universes', M which are completely or almost completely causally disjoint regions of spacetime, within which the parameters of the standard models of physics and cosmology differ. To the extent that they are causally disjoint, we have no ability to make observations in other universe than our own. The parameters of the standard models of particle physics and cosmology vary over the ensemble of universes."* (Smolin (2004) 4)

*"**C** There are many different possible consistent phenomenological descriptions of physics, relevant for the possible description of elementary particle physics at scales less than Planck energies. These may correspond to different phases of the vacuum, or different theories altogether."* (Smolin (2004) 12)

*"**Multiverse hypothesis**. Assuming **A** and **C**, the whole of reality - which we call the multiverse - consists of many different regions of spacetime, within which phenomena are governed by different of these phenomenological descriptions. For simplicity, we call these* universes.*"* (Smolin (2004) 12) unconnected, real worlds means to take for granted (at least for the argument's sake) a very stable realism including the willingness to attribute real ontic status to entities which are empirically inaccessible by definition. So, there is no doubt that the ensemble hypothesis is a metaphysical concept, unprovable empirically on principle. And therefore, there will only be a reasonable motivation for the ensemble hypothesis, if there exists (beside the highly speculative theoretical ideas of string theory) a context of relevance based, at least partially, in empirical evidence (and, furthermore, completely independent of string theory and its landscape hypothesis). But, before turning to such a context (the spectrum of possible explanations for the "fine-tuning" of the universe) it is useful to have a closer look at the ensemble hypothesis and its implications.

Crucial for the transition from the contingency hypothesis to the ensemble hypothesis are the answers to the following questions: What does it mean exactly that a world is not only possible, but also real? Or, that it is possible, but not real? Why is our world not only possible, but also actually existing? Do other possible worlds also actually exist? What does it mean to exist?

## 6. Everything

The most extreme answer to these questions can be found in Max Tegmark's "Ultimate Ensemble Theory":[32]

> *"Physical existence is equivalent to mathematical existence. [...] Mathematical existence is merely freedom from contradiction."* (Tegmark (1998) 7)

The Ultimate Ensemble Theory is in Tegmark's own assessment

> *"[...] a form of radical Platonism, asserting that the mathematical structures in Plato's* realm of ideas*, the* Mindscape *of Rucker, exist 'out there' in a physical sense, akin to what Barrow refers to as 'pi in the sky'."* (Tegmark (1998) 4)

But, as Tegmark suggests, the spectrum of consistent possibilities is probably not as immense as one might think:

> *"Although a rich variety of structures enjoys mathematical existence, the variety is limited by the requirement of self-consistency and by the identification of isomorphic ones."* (Tegmark (1998) 8)

Crucial for Tegmark's Ultimate Ensemble Theory is the question what he means exactly when speaking about the physical existence of consistent mathematical structures. For him the physical existence of a structure means: to appear as physically existing for a "self-aware substructure" the

---

[32] Cf. Tegmark (1998). Tegmark's theory can be seen as the ideal contrasting background to the vacuum selection problem of string theory, at least as long as string theory is supposed not to cover all mathematically consistent possibilities, but only a specific part of these, preselected by specific nomological constraints. It would be most interesting, if it could be shown that string theory already includes all mathematically consistent possibilities, as Wolfgang Lerche suggests:
> *"In view of the many non-trivial consistency constraints that are fulfilled, it is most likely that there is simply no room for a 'different' consistent theory; in other words, it is likely that what we have found is the complete space of all possible consistent quantum theories that include gravity, and string theory may perhaps be viewed as one way of efficiently parametrizing it (in certain regions of its parameter space)."* (Lerche (2000) 33) structure contains. Then, one actually does not need any more the consistency requirement, because inconsistent structures certainly do not contain self-aware substructures.

> *"Our definition of a mathematical structure having [physical existence] was that if it contained a [self-aware substructure], then this [self-aware substructure] would subjectively perceive itself as existing. This means that Hilbert's definition of mathematical existence as self-consistency does not matter for our purposes, since inconsistent systems are too trivial to contain [self-aware substructures] anyway."* (Tegmark (1998) 9)

But, not even every consistent structure will contain self-aware substructures which could perceive the structure they inhabit as existing. And, if a mathematical structure does not contain self-aware substructures, it is, according to Tegmark, completely pointless to ask for the physical existence of the structure.

> *"For the many other mathematical structures that correspond to dead worlds with no [self-aware substructures] there to behold them [...], who cares whether they have [physical existence] or not? [...] The answer to Hawking's question, 'What is it that breathes fire into the equations and makes a Universe for them to describe?' would then be 'you, the [self-aware substructure]'."* (Tegmark (1998) 46)

So, the question of physical existence can be reduced to the question of the existence of self-aware substructures.

> *"We could eliminate the whole notion of [physical existence] from our [theory of everything] by simply rephrasing it as* if a mathematical structure contains a [self-aware substructure], it will perceive itself as existing in a physically real world.*"* (Tegmark (1998) 46)

There are two contrary strategies one can use in order to get scientifically relevant results from Tegmark's Ultimate Ensemble Theory: (i) a top-down approach, starting from the global perspective of the entire spectrum of consistent structures, and (ii) a bottom-up approach, starting from the particular and contingent perspective of a self-aware substructure which does not know which structure it inhabits.[33]

> *"The picture is that some of these mathematical structures contain 'self-aware substructures' [...]. To calculate the physical predictions of the theory, we therefore need to address the following questions:*
> *1. Which structures contain [self-aware substructures]?*
> *2. How do these [self-aware substructures] perceive the structures that they are part of?*
> *3. Given what we perceive, which mathematical structures are most likely to be the ones that we inhabit?"*
> (Tegmark (1998) 39)

In the top-down approach, the objective is to find out which structures exist physically, i.e. which of the consistent structures contain self-aware substructures.

> *"[...] we are asking how large the 'cognizable' part of the grand ensemble is."* (Tegmark (1998) 39)

---

[33] In string theory, the global perspective corresponds to the investigation of the spectrum of vacua within the landscape, and of the statistics of this spectrum; the particular perspective corresponds to the problem to identify, starting from the observable phenomenology, the string vacuum realized in our world.

For that, one has to investigate the entire parameter space of structural possibilities with regard to the question, which constellations lead to the necessary and adequate preconditions for self-aware substructures.[34]

> *"[...] explore the parameter space [...] map out the archipelago of potential habitable islands [...]."* (Tegmark (1998) 48)

The bottom-up strategy corresponds to our usual scientific procedures. Our situation in the world can be characterized, above all, by the fact that it is not at all obvious which structure we inhabit.[35] So, how can this structure be identified from the particular perspective of a self-aware substructure, the observer, the experimentator? This problem is the starting point of all empirical research and, therefore, of the development of scientific theories. It is exactly what makes empirical science necessary. It is the struggle to get from the phenomenology to the nomology, i.e. to the mathematical structure behind appearance. In the case of the structure we inhabit, this investigation is going on at least since Galileo Galilei.[36]

A very interesting feature of the Ultimate Ensemble Theory is that it fulfills the requirements for a fundamental theory in the most radical way: it does not contain any free parameters.

> *"Our [theory of everything] takes this ensemble enlargement to its extreme, and postulates that all structures that exist in the mathematical sense [...] exist in the physical sense as well. The elegance of this theory lies in its extreme simplicity, since it contains neither any free parameters nor any arbitrary assumptions about which of all mathematical equations are assumed to be 'the real ones'."* (Tegmark (1998) 38)

And it does not contain any information at all.[37]

---

[34] Although this is certainly no simple task, Tegmark has already some ideas about the possible results of such an investigation.
> *"However, since the number of constraints for our own particular existence is much greater than the number of free parameters, we argued that it is likely that there is an archipelago of many such small islands, corresponding to different nuclear reaction chains in stellar burning and different chemical compositions of the [self-aware substructures]. The presence of a smaller number of much more severe constraints indicates that this archipelago also has an end, so that large regions on parameter space are likely to be completely devoid of [self-aware substructures], and it is likely that the total number of islands is finite."* (Tegmark (1998) 40)
> *"[...] islands of habitability are small and rare [...] it might even be possible to catalogue all of them."* (Tegmark (1998) 40)

[35] The connecting link between the top-down and the bottom-up approaches consists in the question, how a structure, existing in the Tegmarkian sense, would appear to the self-aware substructures it contains. How would a structure appear from an inner perspective? What phenomenology would a specific structure evoke?

[36] But, according to Tegmark's opinion, there could be limits for our scientific approach. There could be unavoidable ambiguities with regard to the possible structures behind appearance. Different structures might be compatible with our phenomenology. The only viable strategy would then consist in taking the simplest of these structures as the effective nomology leading to observable phenomenology.
> *"Since some aspects of complex mathematical structures can often be approximated by simpler ones, we might never be able to determine precisely which one we are part of. However, if this should turn out to be the case, it clearly will not matter, since we can then obtain all possible physical predictions by just assuming that our structure is the simplest of the candidates."* (Tegmark (1998) 42)

However, the most serious problem is not ambiguity. It is not the case in which we find too many nomological structures compatible with the observable phenomenology; it is rather the case in which we don't find even one empirically adequate nomological structure.
> *"The fact of the matter is that we to date have found no self-consistent mathematical structure that can demonstrably describe both quantum and general relativistic phenomena."* (Tegmark (1998) 48)

[37] Cf. Tegmark (1996)

> *"[...] an entire ensemble is often much simpler than one of its members, which can be stated more formally using the notion of algorithmic information theory [...]. In this sense, our 'ultimate ensemble' of all mathematical structures has virtually no algorithmic complexity at all."* (Tegmark (1998) 44)

On its global level, in the top-down view, the Ultimate Ensemble Theory does away with every contingency. Every alternative theory will contain a certain amount of contingency, if it does not turn out to be identical, finally, to the Ultimate Ensemble Theory. Contingency is in the Ultimate Ensemble Theory restricted to the bottom-up view. But here, it is inevitable. The phenomenology a self-aware substructure can observe within the structure it inhabits is necessarily contingent. Even a maximal bottom-up perspective is restricted by the structure to which it belongs. From the particular perspective, no way leads to a global perspective. The particular perspective is always limited to its structure. So, from the bottom-up perspective, the inexistence of contingency on the global level must always remain a metaphysical hypothesis.

This resembles much the situation in string theory facing the implications of the landscape hypothesis and the vacuum selection problem. Already in the Ultimate Ensemble Theory a mode of argumentation can be seen at work, at least implicitly, which is prevalent within the recent development of string theory and the discussion about the string landscape: It is the idea of anthropic selection and of the so-called "weak anthropic principle".[38]

## 7. Anthropica

The weak anthropic principle itself is nothing but a tautology. It simply reminds us that the world we inhabit must necessarily fulfill conditions that make our existence possible. We can not live in a world which is incompatible with our existence.

> *"This rather tautological (but often overlooked) statement that we have no right to be surprised about things necessary for our existence has been termed* the weak anthropic principle.*"* (Tegmark (1998) 6)

The weak anthropic principle becomes only an interesting tautology in combination with the idea that some or perhaps even most of the possible worlds, more or less different from ours, would probably lead to conditions which make the existence of complex organisms with epistemic capacities impossible. We would never find ourselves in such worlds, simply because we could not live in these worlds. Even if there would be a multitude of worlds, it is evident that we necessarily inhabit a world that has a make-up which allows our existence. This would not be an accidental coincidence, but one of the preconditions of our existence. Not only the principal ontological and nomological possibilities would be decisive for the spectrum of worlds in which we might find ourselves, but, to a far greater amount, the conditions which are indispensable for our existence.

So, an anthropic reasoning with regard to the conditions realized in our world will only have relevance, if there are good arguments supporting the assumption of the existence of an ensemble of worlds, some of which make epistemic subjects possible, and others not. Where could such arguments for an ensemble hypothesis come from? One single speculative theory supporting the ensemble hypothesis (like string theory) does certainly not provide a sufficient motivation for scenarios in which anthropic arguments find their application. There should be additional and independent motivations for the ensemble hypothesis.[39] The best of these independent motivations for the ensemble hypothesis consists in the so-called "fine-tuning" of the universe: There are well-founded indications (based on well-established theories supported by empirical data) that minimal variations of the "natural constants" (coupling parameters, particle masses, cosmological constant etc.) which characterize our world would lead to universes with physical conditions which would make the existence of complex organisms with epistemic capacities very improbable, if not impossible.[40] If life, especially complex organisms with cognitive and epistemic capacities, requires more or less stable and rather complex chemical and cosmic structures, it can be shown to be compatible only with a very small range of parameters: only with minimal variations of the natural constants characterizing our universe. Our universe seems to be made-to-measure for our existence.

---

[38]  Cf. Barrow / Tipler (1986), Hartle (2004), Smolin (2004).

There are only three options to deal with the fine-tuning of the universe: It could be an unexplainable cosmic coincidence. With this option every scientific ambition to explain the fine-tuning ends. If we do not accept the fine-tuning of the universe as the result of pure coincidence, the question remains: How did we get our made-to-measure universe? - It could be a designer-universe, especially and deliberately made for us (or made for other reasons leading to conditions compatible with our existence). This assumption is, if supposing good intentions, the traditional subject of theological considerations and religious belief. If there is no clarity with regard to the intentions, it can also find its place in the context of gnostic scenarios[41] or the recently much discussed "simulation argument"[42] as well as the labyrinthic conceptions of science fiction.[43]

If there remains anything at all for scientific endeavors, it is the third alternative. That our universe is made-to-measure for our existence would be no miracle, if there exists a sufficiently large ensemble of physically real[44] universes. We would find ourselves necessarily in a universe compatible with our existence. The fine-tuning of the universe would be an anthropic selection effect.[45] The ensemble hypothesis given, the fine-tuning of the universe is no surprise.

---

[39] Such independent motivations could even lead to an additional reinforcement of the ensemble hypothesis in the context of string theory.

[40] Cf. Rees (1999), Barrow / Tipler (1986). A minimal variation of the cosmological constant realized in our universe, e.g., would make the formation of large-scale cosmological structures impossible: no galaxies, no formation of second-generation stars and planetary systems containing higher chemical elements, no life. A minimal variation of the parameters of the standard model of elementary particle physics (particle masses, coupling constants etc.) would make atomic structures impossible: no atomic structures, no chemical structures, no life.

[41] Cf. Sloterdijk / Macho (1991).

[42] Cf. Bostrom (2003), Hanson (2001), Schmidhuber (1997), Tipler (1994).

[43] Cf. Galouye (1964). Galouye's novel was adapted by R.W. Fassbinder in his film *Welt am Draht* (1973), remade by J. Rusnak as *The Thirteenth Floor* (1998). *The Matrix* (1999) of the Wachowski Brothers - a gnostic metaphor par excellence - might be at least inspired by these precursors. Cf. Chalmers (n.d.).

[44] If one does not already use Tegmark's terminology, the worlds within this ensemble have to exist. It is not sufficient that these worlds are only possible. See below.

[45] But, one should not think that an anthropical selection would lead to unambiguity with regard to the specific features of the universe in which we find ourselves. It is rather unclear what characteristics exactly are subject to an anthropic selection and how much variation of the relevant parameters is compatible with the existence of intelligent life. Probably, the low energy gauge symmetries, the parameters of the low energy Lagrangian, the matter content (quantity and variety) of the universe, and especially the dimensionality of spacetime will be anthropically decisive and will only admit small variations. The same holds probably for the complexity of material systems which the fundamental nomology admits:

> "Fully linear equations (where all fields are uncoupled) presumably lack the complexity necessary for [self-aware substructures], whereas nonlinearity is notorious for introducing instability and unpredictability (chaos). In other words, it is not implausible that there exists only a small number of possible systems of [partial

However, it is very doubtful, admittedly, that the ensemble hypothesis with its strong realism with regard to, on principle and by definition, inaccessible universes can really be considered a scientific concept. And therefore, it is also very doubtful that the explanation of the fine-tuning of the universe by means of the ensemble hypothesis is a scientific explanation in the strict sense. But the alternatives would be either the assumption of pure cosmic coincidence or the designer-universe scenario. And with these alternatives the scientific endeavor to understand nature ends definitively. So, if anything at all, only the ensemble hypothesis remains.

And with the ensemble hypothesis as an argumentative scenario which, at least, makes the fine-tuning of the universe no miracle, theoretical models and physical theories supporting ensemble scenarios are back in the game. In this context, string theory reenters and possibly even gains additional plausibility.[46] With regard to (i) the fine-tuning of the universe, (ii) the immense multitude of consistent theoretical scenarios resulting from string theory, and (iii) the non-existence of a vacuum selection principle which could be motivated by the theoretical framework, it seems not completely implausible to understand the selection of a string vacuum, representing our world, as an anthropic selection effect.

> *"With nothing preferring one vacuum over another, the anthropic principle comes to the fore whether or not we like the idea."* (Susskind (2003) 17)

However, this would mean that the string scenarios of the landscape are not only nomological possibilities, but that they are physically realized. Only if they are physically realized, an anthropic selection can take place. Contingency alone is not sufficient for anthropic selection. Anthropic selection works only within an ensemble of real worlds. So, a realism with regard to empirically inaccessible worlds seems to be unavoidable. It would only be avoidable, if there were some form of a causal or dynamical connection between the different members of the ensemble.

Recently, Leonard Susskind tried to formulate a concretization of this idea of a dynamical connection by implementing the process of anthropic selection into a cosmological scenario in which the different string vacua are realized by a sequence of dynamical transitions.

> *"To make use of the enormous diversity of environments that string theory is likely to bring with it, we need a dynamical cosmology which, with high probability, will populate one or more regions of space with an anthropically favorable vacuum."* (Susskind (2003) 11)

The idea is that one only needs a sufficiently large number of string vacua with different effective nomology, realized subsequently during cosmic evolution. As a dynamical transition mechanism between the subsequently realized string vacua, vacuum tunneling seems to be the best candidate.[47]

---

> *differential equations] that balance between violating the complexity constraint on one hand and violating the predictability and stability constraints on the other hand."* (Tegmark (1998) 38)

Meanwhile, a sophisticated methodology for the application of anthropic arguments in the natural sciences is under development. Its main objective is to calculate probabilities for the different relevant selectivities. Cf. Aguirre (2005), Stoeger / Ellis / Kirchner (2004), Weinstein (2005).

[46] Other theoretical frameworks, leading to the hypothesis of an ensemble of real worlds, are the scenario of eternal inflation in cosmology and, with certain restrictions, Everett's relative state formulation of quantum mechanics.

[47] According to Susskind, these tunneling processes are unavoidable.

> *"The vacua in string theory with lambda > 0 are not stable and decay on a time scale smaller than the recurrence time."* (Susskind (2003) 17)

*"[...] vacuum tunneling between solutions with different values of the cosmological constant [...] is often assumed to be the mechanism which dynamically implements the anthropic principle. The universe jumps around between vacua until it finds itself in an anthropically allowed one, at which time we observe it."* (Banks / Dine / Gorbatov (2003) 13)

But, should string theory describe nature, and should the ten dimensions of string theory turn out to be a realistic description of spacetime, should furthermore the landscape hypothesis be correct, and should dynamical transitions exist between the scenarios of the landscape, this could lead to rather stormy prospects for our universe.

*"If an observer starts with a large value of the cosmological constant there will be many ways for the causal patch to descend to the supermoduli-space."* (Susskind (2003) 12) *"The potential on the supermoduli-space is zero and so it is always possible to lower the energy by tunneling to a point on the supermoduli-space."* (Susskind (2003) 9) *"The instability also allows the universe to sample all or a large part of the landscape by means of bubble formation."* (Susskind (2003) 17) *"[...] it always ends in an infinite expanding supersymmetric open Fr[i]edman universe."* (Susskind (2003) 15) *"The final and initial states do not have to be four dimensional."* (Susskind (2003) 20)

*"In short [...] if 1) there are extra dimensions of space and 2) the Universe is undergoing accelerated expansion, then the present four-dimensional state of the Universe is not a stable state. The Universe is catastrophically unstable either to decompactification of extra dimensions, to gravitational collapse to a big crunch, or in special cases, possibly to decay to a four-dimensional supersymmetric universe."* (Giddings (2003) 3)

However, for all those who cling to life and need more certainty, there are Banks' already mentioned arguments[48] that, even if a multitude of string scenarios exists, that would not mean that there are dynamical transitions between these scenarios. For the idea of an anthropic selection in string theory this would make impossible any escape route from the strong realism with regard to empirically inaccessible worlds.

## 8. Consequences

Whatever might happen to string theory and to its landscape hypothesis: introducing anthropic reasoning into physics leads to a severe modification of our conception of science and its relation to reality. Not only that the traditional requirements for our most fundamental theories are changing:

---

[48] Cf. Banks (2001), (2003) and (2004). See above. According to Banks, the postulated scenarios forming the landscape are not to be seen as solutions to the same theory, but, if at all, as completely independent theoretical scenarios. But, if the string scenarios within the landscape are not solutions of a common theoretical framework, the idea that there could be dynamical transitions between them is wrong.

*"Unless one rejects the AdS/CFT prescription for quantum gravity in Anti de Sitter space, it is difficult to defend the idea that there is a unique theory of quantum gravity, with different realizations of it corresponding to minima of an effective potential. This field theory inspired picture is based on a separation between UV and IR physics which is simply not there in theories of quantum gravity. I have tried to investigate both real and virtual transitions between vacua with different values of the cosmological constant, or isolated vacua with the same values of the cosmological constant and found that they do not occur - black holes get in the way."* (Banks (2003) 76)

And without the possibility of dynamical transitions between the string vacua there would be no dynamical vacuum selection.

*"[...] in an anthropic theory simplicity and elegance are not considerations. The only criteria for choosing a vacuum is utility, i.e. does it have the necessary elements such as galaxy formation and complex chemistry that are needed for life."* (Susskind (2003) 5f)

Anthropic reasoning touches one of the traditionally central goals of physics. As Steven Weinberg wrote long ago:

*"After all, we do not want merely to describe the world as we find it, but to explain to the greatest possible extent why it has to be the way it is."* (Weinberg (1977) 34)

If an ensemble scenario, without an intrinsic, intratheoretically motivated selection principle, leading to the idea of anthropic selection, should turn out to be unavoidable, this would probably be the end of a purely physical answer to Weinberg's question, why the world is as it is. We would possibly be able to find out to a certain degree how the world is. But the way it is would not be exclusively subject to a physical, but to a cosmological or to an evolutionary explanation, reflecting its contingency. - But this might depict things exactly as they are. If we are not too fond of the metaphysical ideals of unambiguity and uniqueness, and if we are, at the same time, prepared to embrace a rather unlimited metaphysical realism with regard to possible worlds, we might live very well with the contingency of the world.

However, even if the landscape hypothesis of string theory should not turn out to be an artifact, resulting from an abstruse theoretical dead end, completely detached from any empirical control: does the inclusion of anthropic selection really help us to decide, if string theory is an adequate fundamental description of nature? - Even if we would know much more about anthropically suitable conditions,[49] anthropic reasoning would never lead to unambiguous results; it leads to an only very limited predictability. If the landscape hypothesis should turn out to be a necessary implication of string theory, we would only achieve unambiguous results, if it could be shown that none of the scenarios of the string landscape are anthropically suitable, possibly because all scenarios would necessarily have unbroken supersymmetry. This would falsify string theory. It would be our existence that falsifies string theory.

But, if the landscape hypothesis turns out to be a necessary implication of string theory, this is very probably the only way in which a falsification of string theory will be achievable. Otherwise it can not be falsified, because, as Lee Smolin points out, the idea that the selection of a vacuum from the string landscape should be understood as an anthropic effect leads very easily to the following self-immunization:

---

[49] As long as it is not possible to derive the concrete low energy implications from the theory, it is simply undecidable, anyway, if a certain scenario is anthropically suitable. But, even if the low energy physics could be derived for all relevant scenarios, it would probably remain undecidable to a certain degree which scenarios permit complex organisms with epistemic potential, because the conditions for intelligent life are to a large degree unknown.
  *"[...] I do not know how to implement the anthropic principle. It is nearly impossible to say: the weak anthropic principle (the requirement that we find ourselves in an environment or neighborhood which can support life) requires the cosmological constant to be..., the fine structure constant to be..., the strength of inflationary fluctuation to be... The problem is simply too complicated."* (Dine (2004a) 8)
  *"It is likely that the low energy dynamics of any theory satisfying our criteria would be sufficiently complicated that we would have little chance of deciding whether complex, intelligent organisms could evolve in these alternative universes."* (Banks (2003) 55)

*"'Our theory has many solutions $S_i$. One of them, $S_1$ gives rise to a prediction X. If X is found that will confirm the combination of our theory and the particular solution $S_1$. But if X is not found belief in the theory is not diminished, for there are a large number of solutions that don't predict X.'"* (Smolin (2004) 5)

Has string theory, with the landscape hypothesis and with its understanding of the selection as an anthropic effect, already reached the stage of this self-immunization?

*"It is indeed plausible that this is already the case with string theory [...]."* (Smolin (2004) 5)

If anthropic selectivity dominates completely the relation between a theory and the observable phenomenology, one obviously does not have to know much about the fundamental theory, except that it includes, anyhow, with a probability above zero, anthropically suitable solutions.

*"Perhaps the most attractive feature of the anthropic argument is that it does not require us to know much about the Meta-theory, which determines the probability distribution of the cosmological constant. One requires only that such a theory exists and that the probability distribution in the vicinity of the anthropic bound is nonzero, and reasonably smooth. The lack of dependence on details of the Meta-theory is important, because it is unlikely that any of those details could be checked by experiment. If we needed to understand an elaborate mathematical theory, most of whose structure could never be tested, in order to believe in the anthropic bound, then that bound would appear much less plausible."* (Banks (2003) 46f)

Given a theory with anthropically suitable solutions, a theory which could describe the fundamental nomology of our world, the suspicion always remains, as long as there are no independent empirical tests, that there could be other theories with the same virtues, but postulating completely different fundamental nomologies.

*"I must admit to a great deal of unease in talking about these arguments. Consider the following model of a Meta-theory: A supreme being plays dice with himself, and on the basis of each throw, decides to construct a universe with a finite number of quantum states obeying the famous, yet to be constructed, rules for quantum cosmology in such a universe. Only the number of spins n is decided by the throw of the dice. We then apply the anthropic argument. As theoretical physicists, we would certainly find an elegant mathematical model of a Meta-theory more satisfying than the supreme being model. But our inability to perform experiments for the values of n that are ruled out by the anthropic argument, leaves us with no experimental proof that the supreme being model is any less right than the mathematical one. We must ask ourselves whether we are really doing science. So must anyone who indulges in anthropic speculation."* (Banks (2003) 48)

How could we ever trust in such a theory as a fundamental description of nature? Ensemble theories whose relevance for the observable phenomenology is mediated exclusively by anthropic selectivity can not really be falsified. This is completely unacceptable for a scientific theory:

*"The simple reason is that once a non-falsifiable theory is preferred to falsifiable alternatives, the process of science stops and further increases in knowledge are ruled out."* (Smolin (2004) 5)

So, are there alternatives to anthropic reasoning with regard to the selection of scenarios from the string landscape, or to the string landscape itself, or even to string theory? Or is this simply the end?